\documentclass{iau} 
\usepackage{graphicx}

\title[Measurements of the global 21-cm signal from the Cosmic Dawn] 
{Measurements of the global 21-cm signal from the Cosmic Dawn}

\author[Gianni Bernardi]   
{Gianni Bernardi$^{1,2,3}$
}

\affiliation{$^1$INAF-Istituto di Radioastronomia, via Gobetti 101, 40129, Bologna, Italy \\ email: {gbernardi@ira.inaf.it} \\[\affilskip]
$^2$Department of Physics and Electronics, Rhodes University, PO Box 94, Grahamstown, 6140, South Africa\\
$^{3}$SKA SA, 3rd Floor, The Park, Park Road, Pinelands, 7405, South Africa}

\pubyear{2018}
\volume{333}  
\jname{Peering towards Cosmic Dawn}
\editors{Vibor Jeli\'c \& Thijs van der Hulst, eds.}
\begin{document}

\maketitle

\begin{abstract}
The sky-averaged (global) 21-cm signal is a very promising probe of the Cosmic Dawn, when the first luminous sources were formed and started to shine in a substantially neutral intergalactic medium. I here report on the status and early result of the Large-Aperture Experiment to Detect the Dark Age that focuses on observations of the global 21-cm signal in the $16 \lesssim z \lesssim 30$ range.
\keywords{cosmology:observations - dark ages, reionization, first stars - diffuse radiation.}
\end{abstract}

\firstsection 
\section{Introduction}

Observations of the highly-redshifted 21-cm emission are considered the most powerful probe of the birth of the first luminous sources and
the consequent epoch of reionization (see \cite{mcquinn16}; \cite{furlanetto16}; for recent reviews). In particular, the sensitivity required to measure the sky--averaged (global) 21-cm signal may be achieved with relatively short observations (from a few tens to a few hundreds of hours) using a single dipole. Before widespread reionization, observations of the global 21-cm signal would target two specific transition phases:
\begin{itemize}
\item the birth of the first luminous sources in the Universe, expected to happen at $z \sim 25-30$ and to tightly couple the spin temperature to the InterGalactic Medium (IGM) temperature, generating 21-cm emission via resonant scattering of Ly$\alpha$ photons (\cite{field59}). As the IGM is colder than the Cosmic Microwave Background (CMB) at these epochs, the 21-cm is expected to appear in absorption against the CMB. The intensity and timing of this absorption feature depends strongly on the intensity of the Ly$\alpha$ radiation from the first stars as well as the feedback processes occurring in these first galactic halos (e.g., \cite{furlanetto06});
\item the IGM heating epoch, when X-rays from the first stellar black holes starts to heat the IGM, eventually driving the gas temperature above the CMB (e.g., \cite{pritchard07}; \cite{mesinger13}).
\end{itemize}
The Large-Aperture Experiment to Detect the Dark Age (LEDA; \cite{greenhill12}; \cite{bernardi15}) aims to detect the global 21-cm signal in the $16 \lesssim z \lesssim 30$ range, attempting to characterize the birth of the first luminous sources as well as the X-ray heating epoch.

\section{LEDA observations of the global 21-cm signal from the Cosmic Dawn}

\begin{figure}
\centering
\includegraphics[width=1.\columnwidth]{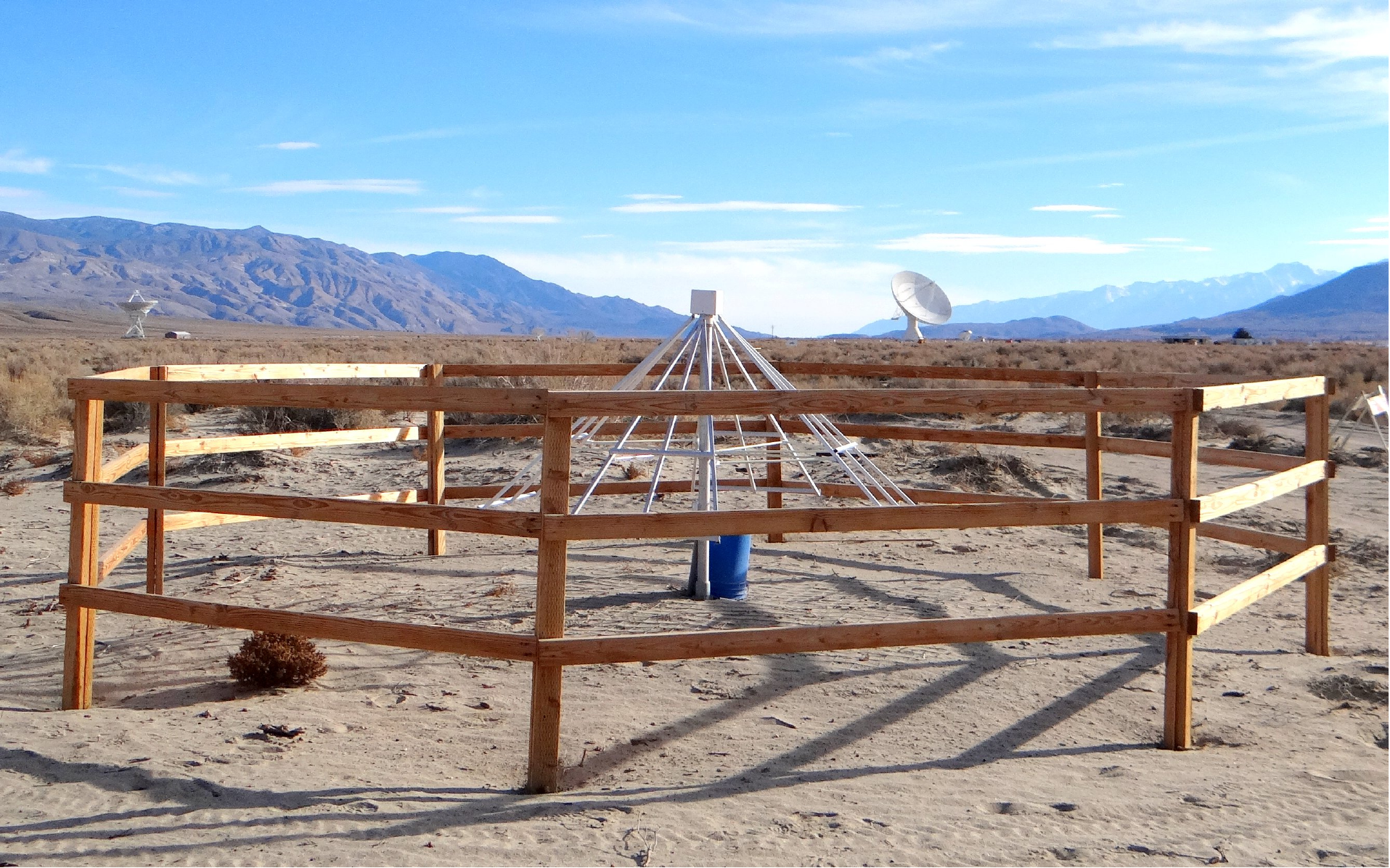}
\caption{One of the LEDA outrigger antenna stands equipped for global 21-cm observations and located at the Owens Valley Radio Observatory (from \cite{price17}).}\label{fig:LEDA_outrigger} 
\end{figure}
The biggest challenge that global 21-cm observations need to face is the separation of bright foreground emission that, for single dipole experiments, is essentially Galactic synchrotron radiation. The separation leverages on the spectral difference between the 21-cm signal and the synchrotron foreground and requires to achieve a high accuracy in both radiometric calibration (\cite{monsalve17}) and accurate measurement (and/or mitigation) of the chromatic instrument primary beam (\cite{bernardi15}; \cite{mozdzen17}).   

LEDA (Figure~\ref{fig:LEDA_outrigger}) adopts a twofold strategy in order to mitigate systematic effects:
\begin{itemize}
\item the development of custom electronics in order to achieve accurate calibration of radio frequency signal chain over a wide bandwidth (\cite{price17}). The custom-developed receivers are deployed on four dipoles so that independent measurements help to further identify and correct systematic effects;
\item cross-correlation of the individual dipoles with an array of $\sim 250$ similar dipoles spread over a $\sim 200$~m diameter area to provide an interferometric array capable of observing at frequencies between 30 and 88~MHz. Interferometric techniques can be used to measure the spatial and spectral structure of the dipole beam (\cite{bernardi15}).
\end{itemize}

Early observations were carried out on February $11^{\rm th}$ 2016. Two hours of data were calibrated using a three-position switch in order to correct for time variations of the receiver gains and set the absolute flux density scale. The final calibrated sky temperature was achieved by correcting for frequency structure in the bandpass using both lab measurements and models of the synchrotron emission and the dipole beam response (further details of the data calibration can be found in \cite{bernardi16}). 

\begin{figure}
\centering
\includegraphics[width=1.\columnwidth]{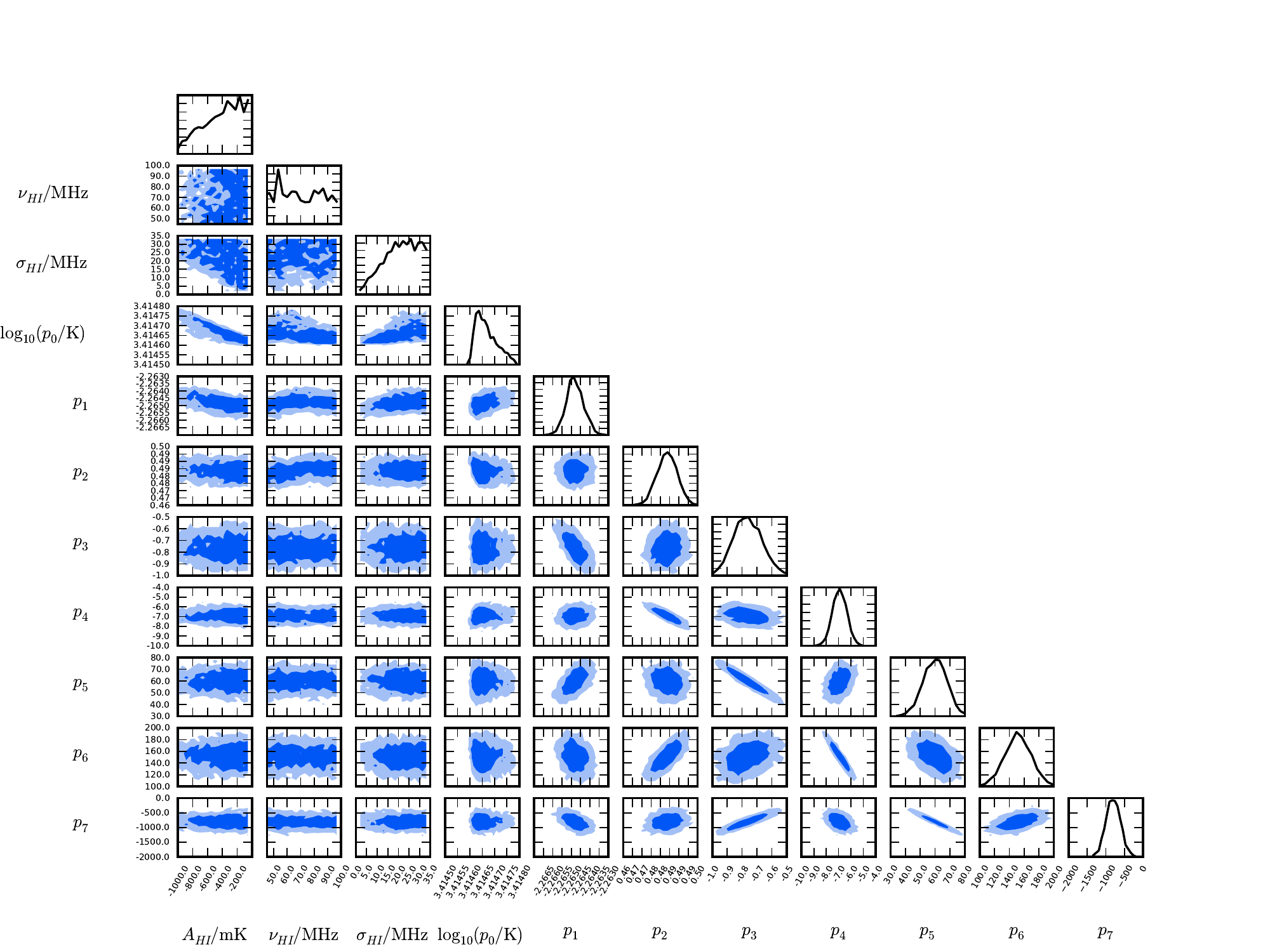}
\caption{Posterior probability distribution, marginalized into one and two dimensions and plotted in the $[0,1]$ range, for the synchrotron foreground and 21-cm models, fitted to the LEDA data. The dark and light shaded regions indicate the 68- and 95-per-cent confidence regions (from \cite{bernardi16}).}\label{fig:triangle-plot} 
\end{figure}

Foreground separation was perfomed using a Bayesian sampling of the likelihood function (\cite{harker12}; \cite{bernardi16}):
\begin{equation}
{\mathcal L}_j\left(T_{\rm ant} (\nu_j) | {\bf \Theta}\right) = \frac{1} {\sqrt{2 \pi \sigma^2(\nu_j)}} \mathrm{e}^{-\frac{[T_{\rm ant}(\nu_j) - T_m(\nu_j,\mathbf{\Theta})]^2}{2 \sigma^2(\nu_j)}},
\end{equation}
where $T_{\rm ant}(\nu_j)$ is the observed sky temperature at the frequency $\nu_j$, $T_m(\nu_j,\mathbf{\Theta})$ is the model spectrum function of the parameter set $\mathbf{\Theta}$, and $\sigma(\nu_j)$ is the standard deviation of the instrumental noise. The model spectrum was taken as the sum of a log-polynomial expansion for the synchrotron spectrum and a Gaussian absorption trough for the 21-cm signal:
\begin{eqnarray}
T_m (\nu_j,p_0,...,p_7,A_{\rm HI},\nu_{\rm HI},\sigma_{\rm HI}) = 10^{\sum_{n=0}^7 p_n \left[ \log_{10} \left( \frac{\nu_j}{\nu_0} \right) \right]^n} +
A_{\rm HI} \, \mathrm{e}^{-\frac{(\nu_j - \nu_{\rm HI})^2}{2 \sigma^2_{\rm HI}}}.
\end{eqnarray}
Figure~\ref{fig:triangle-plot} shows the recovered best fit parameters. Measurements led to fairly tight constraints on the foreground parameter space and to the first data-driven constraints on the 21-cm signal parameters, leading to reject Gaussian like models with amplitude $A_{\rm HI} > -890$~mK {\it and} width $\sigma_{\rm HI} > 6.5$~MHz at the 95-per-cent confidence level in the $16 \lesssim z \lesssim 30$ range.


\section{Conclusions}

I have presented an update on the search of the global 21-cm emission in $16 \lesssim z \lesssim 30$ from LEDA, focused to understand the formation of the first sources in the Universe. Initial data place constraints on the signal amplitude below 890~mK in absorption against the CMB and future longer observations may be able to place maningful constraints on extreme 21-cm models for which, for example, structure formation starts at lower redshift with a very high efficiency. 

\section*{Acknowledgements}
This work is based on the research supported in part by the National Research Foundation of South Africa (grant No. 103424).

\bibliography{Bibliography}





\end{document}